\documentclass[baaa]{baaa}

\usepackage[pdftex]{hyperref}
\usepackage{subfigure}
\usepackage{natbib}
\usepackage{helvet,soul}
\usepackage[font=small]{caption}

\begin{document}


\journalvol{60}
\journalyear{2018}
\journaleditors{P. Benaglia, A.C. Rovero, R. Gamen \& M. Lares}


\contriblanguage{1}


\contribtype{3}

\thematicarea{6}

\title{The interstellar medium: from molecules to star formation}


\titlerunning{The interstellar medium}


\author{S. Paron\inst{1}}
\authorrunning{Paron et al.}


\contact{sparon@iafe.uba.ar}

\institute{CONICET-Universidad de Buenos Aires. Instituto de Astronom\'{\i}a y F\'{\i}sica del Espacio. 
}


\resumen{
El medio interestelar (MIE) es un medio extremadamente complejo que contiene toda la materia
necesaria para la formación de estrellas y sistemas planetarios. El MIE se encuentra en permanente
interacción con la radiación, turbulencia, campos magnéticos y gravitatorios, y partículas aceleradas.
Todo lo que ocurre en este medio tiene consecuencias en la dinámica y evolución de la galaxia, resultando
así ser el enlace que relaciona las escalas estelares con las galácticas.
El estudio del MIE es entonces fundamental para avanzar de una manera integral en el conocimiento de la astrofísica
estelar y galáctica.
En este trabajo se presenta un resumen de lo que conocemos hoy en día sobre la física y química de este medio,
haciendo especial énfasis en la formación de estrellas, y cómo éstas a través de los procesos relacionados
a su nacimiento y evolución, se interrelacionan con el medio que las rodea y contiene.
}

\abstract{
The interstellar medium (ISM) is a very complex medium which contains the matter needed to 
form stars and planets. The ISM is in permanent interaction with radiation, turbulence, magnetic and gravitational fields, 
and accelerated particles.
Everything that happens in this medium has consequences on the dynamics and evolution of the Galaxy, resulting the link that relates 
the stellar scale with the galactic one. Thus, the study of the ISM is crucial to advance in the knowledge of stellar and galactic astrophysics.
In this article I present a summary of what we know about the physics and chemistry of this medium, giving an special emphasis 
on star formation, and how the processes related to the stars birth and evolution interrelate with the environment 
that surrounds them.

}



\maketitle

\section{Defining and classifying the interstellar medium}
\label{def}
 
All in the Galaxy is embedded in the interstellar medium (ISM), which is a dynamic, ever-changing environment. 
We can say that the Galaxy, besides the dark matter, is composed by stars and regular matter distributed between 
them, which conforms a medium of low density. The ISM is not homogeneous, indeed it can have, from place to place, very different physical 
and chemical conditions according to the influence of the radiation, turbulence, 
magnetic fields, and cosmic rays. In that sense it is not possible to classify the ISM among well defined 
components, and thus, in the literature it is 
usually given schematic classifications, such as in \citet{leq05} (see Table\,\ref{tism}), or in \citet{tielens05}, between
others. 

Table\,\ref{tism} shows that all the ISM phases or components are extremely tenuous comparing with terrestrial environments.
For instance, if we consider
even the densest phase, i.e. the molecular phase, which can have densities from some $10^{3}$ to about $10^{7}$ cm$^{-3}$, this is
several orders of magnitude less dense than, for example, the atmosphere of our planet that contains about $10^{19}$ molecules per cm$^{3}$.

\begin{table}[!h]
\centering
\caption{Schematic classification for the components of the ISM from \citet{leq05}.}
\tiny
\begin{tabular}{lllll}
\hline\hline\noalign{\smallskip}
Phase      &             &  Density         & Temperature  \\
           &             &  (cm$^{-3}$)     &  (K)         \\   
\hline\noalign{\smallskip}
Atomic (HI)          &  Cold                  &  $\simeq$ 25               & $\simeq$ 100              \\
                     &  Warm                  &  $\simeq$ 0.25             & $\simeq$ $8\times10^{3}$   \\
Molecular (H$_{2}$)  &                        &  $\geq 10^{3}$             & $\leq$   100                \\
Ionized              &  H{\sc ii} regions     &  $\simeq$ 1 -- $10^{4}$    & $\simeq  10^{4}$             \\
                     &  Diffuse               &  $\simeq$ 0.03             & $\simeq   8\times10^{3}$      \\
                     &  Hot                   &  $\simeq 6\times10^{-3}$   & $\simeq   5\times10^{5}$       \\
\hline
\end{tabular}
\label{tism}
\end{table}

The matter of the ISM represents about 5\%~of the total stellar mass in the Galaxy, and its composition is about 
70\%~hydrogen, 28\%~helium, and 2\%~of heavy elements \citep{leq05}. In turn, these elements can be neutral or ionized,
and can be assambed into molecules. From the total matter of the ISM, the fraction of atomic ionized matter is 
23\%, while the atomic neutral matter represents a 60\%. Finally, the portion of molecular matter is about 17\%~\citep{draine11}.

In spite of being a medium of low density and representing an small fraction of the total mass in the Galaxy, the ISM is extremely important
for the galactic evolution. It contains all the matter to form new generations of stars and planets, and its molecular phase is composed
by complex molecules, including pre-biotic ones. Thus, this article is focused on the molecular component of the ISM and the star formation 
processes that occur in it.

\section{Molecules in the ISM}
\label{mol}

Although astronomical molecules are detected in diverse astrophysical environments, the molecular material in general, and the complex
molecules in particular, are mainly associated with molecular clouds and their dense clumps, which comprises both a gaseous
phase and a solid one of small dust particles. The analysis of the molecular spectra and the chemical chains of reactions that yield  
different molecular species are useful to probe the ISM components and the processes that occur in it. The
molecular spectra gives physical information on the gas and dust. In particular, 
rotational and vibrational spectra tell us about the density and temperature of the gas as well as the kinematics of it, 
such as collapse and rotation \citep{herbst09}.

The richness of the molecular component of the ISM is evidenced by the large number of known molecules. Up to date around 200 molecules 
have been detected in the ISM or circumstellar shells. For a quick revision of the known molecules I recommend to visit 
https://www.astro.uni-koeln.de/cdms/molecules from the University of Cologne, where it can be find an up to
date list of molecules ordered by the number of atoms that compose them. Another interesting website where it is presented
the known molecules ordered as chronologically as were discovered is: http://www.astrochymist.org/astrochymist\_ism.html  
from the site http://www.astrochymist.org/, where there are much information and links about astrochemistry.
As can be seen from these catalogues and from many works in the literature (see for example \citealt{ehren00,herbst09,bell13,bonfand17}), 
a large number of the known molecular species are complex and organic molecules.

\subsection{Complex molecules: the role of the dust}

Given to the low densities, the intense UV radiation and the constant flux of cosmic rays, 
the main chemical reactions in the ISM should be quite different than those that take place in our planet. Nowadays  
it is well established that the main chemical reactions in the ISM are the ion-molecule reactions, and the chemistry on the surface of the
interstellar dust. The first one is part of the gas-phase chemistry, which also include others reactions such as radiative association, 
dissociative recombination, neutral-neutral reaction, and photodissociation and photoionization. For a review of these chemical
processes, and their relative importance, see for example \citet{leq91}, \citet{herbst95}, and \citet{leq05}. 
A more recent overview of astrochemistry with extensive reference lists can be found in \citet{tielens13}.
Here, I focus on the role of the interstellar dust in the formation of complex molecules in dark clouds, the sites of star formation. 

\begin{figure}[!h]
  \centering
  \includegraphics[width=0.46\textwidth]{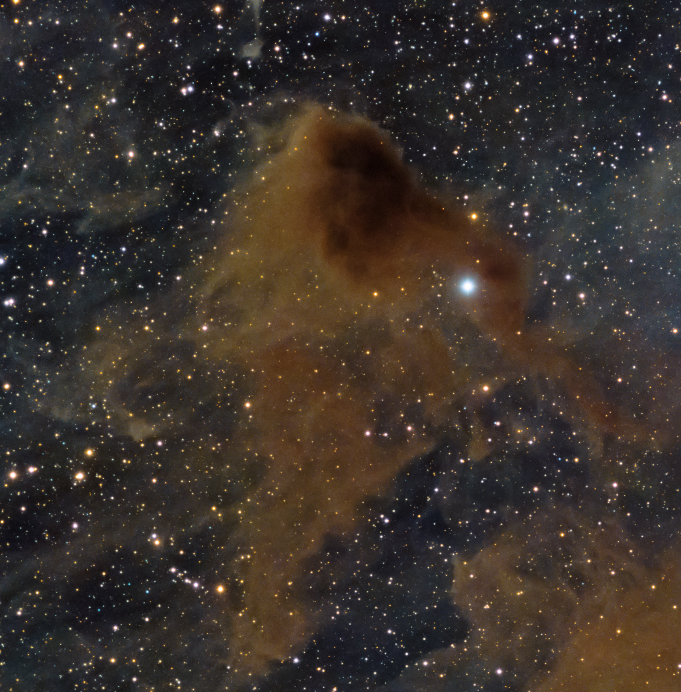}
  \caption{Beverly Lynds Dark Nebula 183. This is an example of the visual extinction due to
the presence of interstellar dust. Image extracted from https://apod.nasa.gov/apod/ap171021.html; credit \& copyright:
Fabian Neyer.}
  \label{dark}
\end{figure}

\begin{figure}[!h]
  \centering
  \includegraphics[width=0.22\textwidth]{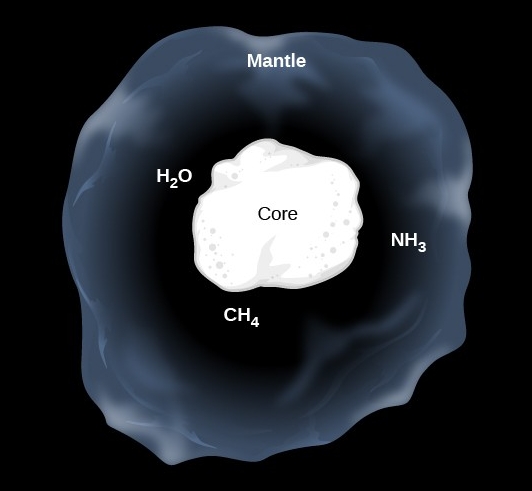}
  \includegraphics[width=0.26\textwidth]{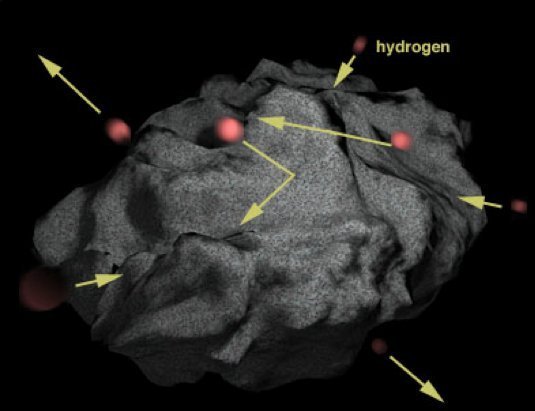}
  \caption{Left: Cartoon of a dust grain with an icy molecular mantle (image extracted from 
the cosmic-dust chapter in https://courses.lumenlearning.com/astronomy/). Right: Cartoon of the interaction of hydrogen atoms with
a dust grain  (image from \citet{vidali13a}).}
  \label{dust}
\end{figure}

The interstellar dust is the main component of the ISM responsible for the visual extinction (Figure\,\ref{dark}) and the study
of its composition began about a century ago. For instance, \citet{spitz41} proposed, not so wrongly, the presence of metallic 
particles embedded in the clouds distributed in the galactic plane. Along the years much effort was done in order to explain the composition of 
this solid component. In this 
context, for example \citet{gould63}, following \citet{vanhulst49}, considered dust grains that consist of something like between a 
hydrogen-bonded crystal (ice) and a molecular crystal, while \citet{stecher66} mentioned
grains of pure-graphite. To mention a flamboyant case, \citet{hoylea,hoyleb} suggested that microbiological systems such as virus, bacteria,
and algae could explain the interstellar extinction.
Nowadays we know that the solid component of the ISM is composed by tiny solid particles of $\sim0.01-0.5~\mu$m 
in size consisting of silicates and carbonaceous material \citep{draine03,draine07}, that in the cold dense clouds the gas-phase chemical 
species condense on them forming icy mantles (see Fig.\,\ref{dust} left).

The interstellar dust is responsible of the formation of the most basic and abundant molecule in the Universe, the molecular hydrogen (H$_{2}$).
Roughly, the principle of H$_{2}$ formation is simple: two H atoms stick onto a dust grain, encounter one another and form
the molecule \citep{hollen71}. This stickiness process can be due to physical and/or chemical adsorption (e.g. \citealt{vidali13b}),
which depend on the dust temperature, composition, and presence, or not, of a mantle. Once on the surface, the H atom can
evaporate or hop across it until encountering another H atom and form the molecule (Fig.\,\ref{dust} right). 
Finally, H$_{2}$ stays on the surface until a temperature is reached that allows its evaporation, injecting the simplest molecule to 
the gas-phase of the ISM.

\begin{figure}[!h]
  \centering
  \includegraphics[width=0.5\textwidth]{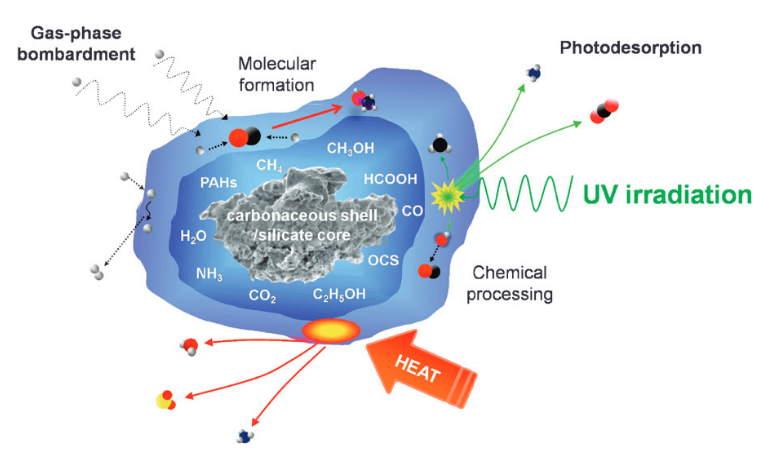}
  \caption{Schematic illustration of some processes that take place on icy grains. Image from \citet{burke10}.}
  \label{dustchem}
\end{figure}

The role of the interstellar dust in the most complex chemistry that take place in the ISM is crucial. This is because they
contribute to enrich the gas-phase with simple and quite complex molecular species formed onto them, and shield regions from the
UV radiation, protecting in some cases large chemical chains. Indeed many molecules can form on grains. Some can form inside the 
mantles, and others on the grain surface, which can take part
in further surface chemical reactions. The dust chemistry is in permanent association with the gas-phase chemistry \citep{hocuk15}. The 
molecular species formed in both the mantles and the dust surface can be injected into the gas-phase through evaporation due to the absorption
of UV photons (photo-desorption), or heating (thermal-desorption), or by escaping after total or partial 
grain disruption produced by collisions (see e.g. \citealt{dhend85,biss07,miura17}). Figure\,\ref{dustchem} illustrates some of these
processes. 

Since the stars form in dark and cool regions plenty of interstellar dust which, as we see, favours the complex chemistry, 
understanding the chemical evolution of star-forming regions is a central topic. By one 
side, it is important because the chemistry is present along the whole process of stars and planets formation, which, moreover 
can lead to the biology; and on the other side, the chemistry can be used to probe the physical conditions of such 
regions \citep{gerner14}. 

Next Section focuses on star formation processes in the context of the ISM as a dynamic and 
ever-changing environment.

\section{Star formation}

Taking into account that all in the ISM is in interaction, it is not possible to think the star formation as an isolated process. 
This is the reason why since some decades the so-called triggered star formation mechanisms are being  
investigated, mainly in the case of the formation of massive stars, in which one, 
or several events compress the molecular gas and initiate its collapse. In this Section, I review some and show the usual
procedures to investigate them.

\subsection{Collect and collapse} 

One of the star forming triggered processes that has been widely studied around H{\sc ii} regions is the so called ``collect and collapse''
(CC).
It was early proposed by \citet{elme77} and numerically treated by \citet{whit94a,whit94b}.
In such process, during the supersonic expansion of an H{\sc ii} region, a dense layer of material can be collected
between the ionization and the shock fronts. Then this layer can be fragmented in
massive condensations that then can collapse to lead the formation of new massive stars and/or clusters.

\begin{figure}[!h]
  \centering
  \includegraphics[width=0.45\textwidth]{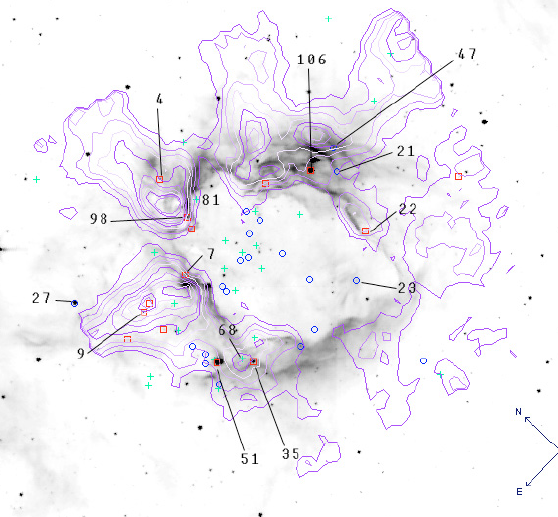}
  \caption{Example of a molecular shell surrounding an H{\sc ii} region with presence of YSOs. The image was extracted
from \citet{poma09}. The borders of the H{\sc ii} region are traced with the 8 $\mu$m emission (in gray)
and the molecular shell, shown in contours, is mapped through the emission of $^{13}$CO. Red squares are the positions of
Class\,I sources, i.e. the youngest YSOs.  }
  \label{cc}
\end{figure}

To investigate whether the CC process is ongoing, firstly, using molecular and/or dust emission, 
it is looking for a dense molecular shell surrounding the ionized gas of an H{\sc ii} region, or molecular massive fragments 
regularly spaced along the ionization front. Then, it is necessary to perform a search and analysis of point sources related to
the molecular shell or fragments in order to determine if they are young stellar objects (YSOs).
Finally, if a molecular shell and associated YSOs are found surrounding the H{\sc ii} (see Fig.\,\ref{cc}), 
in order to establish whether the CC mechanism is responsible of that, a comparison 
between the dynamical age of the H{\sc ii} region and the fragmentation 
time predicted by the theoretical models of Whitworth et al. must be done. This kind of investigations can be found
in many works in the literature, see for example \citet{deha09,poma09,zav10,duro17}. 

It is important to note that several processes can trigger star formation in the vicinity of H{\sc ii} regions, and
the mere finding of a molecular shell surrounding an H{\sc ii} region not should be interpreted as a conclusive probe of
the CC mechanism. In fact, this process can be acting in combination with others
(e.g. \citealt{walch15,deha15}) such as the radiative driven implosion, which is discussed in the next Section.

\subsection{Radiative driven implosion}

\begin{figure}[!h]
  \centering
  \includegraphics[width=0.45\textwidth]{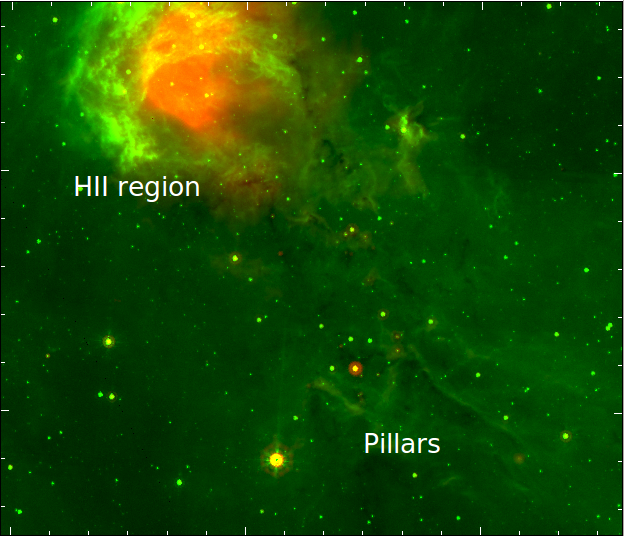}
  \caption{Example of an H{\sc ii} region irradiating pillar-like shape molecular condensations. The 8 and 24 $\mu$m emissions
are displayed in green and red, respectively. From \citet{paron17}. }
  \label{pils}
\end{figure}

Another triggered mechanism of star formation that may occur around H{\sc ii} regions is the radiation-driven implosion (RDI),
first proposed by \citet{reip83}. This process can begin when the ionization front from the H{\sc ii} region moves over a molecular 
condensation, which generates a dense outer shell of ionized gas, usually called the ionized boundary layer (IBL). 
If the IBL is over-pressured with respect to the interior of the molecular condensation,
shocks are driven into it compressing the molecular material until the internal pressure is balanced with the pressure of
the IBL. At this stage,  clumps within the compressed molecular feature can collapse leading to the creation of a new generation
of stars \citep{berto90,leflo94}.
Recent theoretical and observational studies of pillars, molecular features that are related to H{\sc ii} regions, and possible 
star formation linked to them can be found in \citet{tremblin12,tremblin13} and \citet{hartigan15}. Figure\,\ref{pils} shows 
pillar-like structures pointing to the open border of an H{\sc ii} region, suggesting that they were sculpted by the 
ionization flux and hence are suitable features to investigate RDI.
This kind of scenarios are compatible with the theoretical models developed by \citet{grit09b,grit09a}, from which they
show that the formation of pillar-like structures can be explained by the radiation from hot stars penetrating
through low density interclump channels in the molecular clouds. In addition, these models also predict the formation
of a new generation of stars in the pillars heads.

To investigate such a process it is necessary to estimate the ionized and neutral gas densities at the IBL and at the interior of the
radiation exposed molecular structure, respectively. The first one can be obtained quite directly from, for example, observations of 
the radio continuum 
emission from the ionized gas at the IBL, or indirectly based on, knowing the spectral type of the H{\sc ii} 
region exciting star(s), 
an estimate of the amount of UV photons arriving to the molecular structure. In the case of the neutral gas density at the interior
of the irradiated molecular feature, it can be derived directly from molecular observations.
Once one has the densities values, external and internal pressures can be derived in order to compare them and evaluate whether the 
RDI process is at work (e.g. \citealt{thomp04,ortega13}). On the other side, \citet{bisbas11} show, in numerical simulations, that it 
is also possible studying the RDI process only evaluating the incident ionizing flux over the molecular structure.  

Additionally, this kind of molecular structures exposed to the UV ionization from an H{\sc ii} region are perfect sites to probe
the ion-molecule reactions. 
For instance, the detection of molecular species such as HCN, HNC, H$_{3}$O$^{+}$, HCO$^{+}$, between others,
towards the tips of the irradiated molecular structures allows us to study chemical chains in which the radiation 
has an important role and to probe the influence of the UV irradiation in the dense gas \citep{grani14,chenel16,paron17}. 

\subsection{Cloud-cloud collisions}

\begin{figure}[!h]
  \centering
  \includegraphics[width=0.45\textwidth]{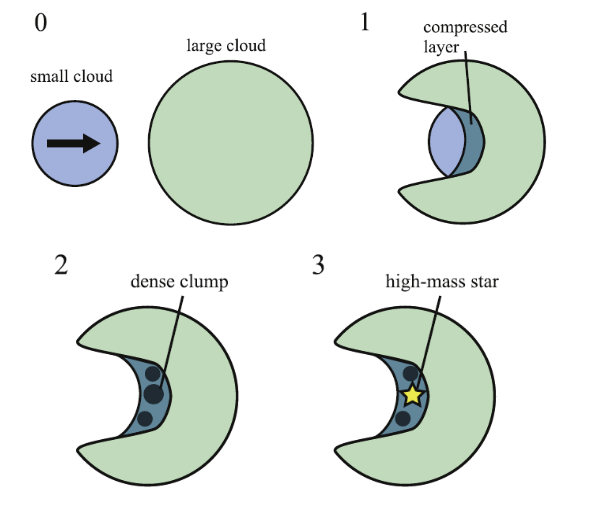}
  \caption{Schematic illustration of triggered star formation due to cloud-cloud collisions as studied in \citet{habe92}. 
Image from \citet{torii17}. }
  \label{coll}
\end{figure}

Taking into account the dynamics of the ISM, collisions between molecular clouds could happen frequently. 
Cloud-cloud collisions were proposed as an another important triggered star formation process. Early observational 
evidence of such a mechanism was presented by \citet{loren76}, and later it was numerically studied by \citet{habe92}. 
The cloud-cloud collision generates a dense gas layer at the interface of the collision where it can be induced 
the formation of dense self-gravitating clumps (Fig.\,\ref{coll}). At this interface the magnetic field and the 
gas turbulent velocity amplify, which increase the mass accretion rate of the clumps leading to the formation of high mass stars
\citep{inoue13}.

\begin{figure}[!h]
  \centering
  \includegraphics[width=0.49\textwidth]{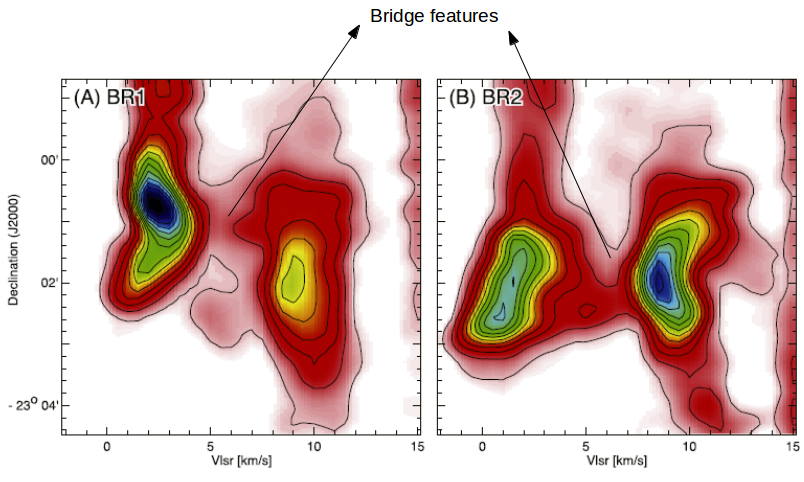}
  \caption{Position-velocity diagrams of the $^{12}$CO J=3--2 emission towards a region where two clouds are likely colliding.
Image from \citet{torii17}. }
  \label{brid}
\end{figure}

To perform observational studies of this mechanism, signatures of collision must be detected using molecular lines observations,
and young massive stellar objects related to the molecular structure have to be observed embedded in the junction of the colliding
clouds.
One important signature characteristic of the collision between two dissimilar clouds is
the cavity generated on the larger cloud (see Fig.\,\ref{coll}). Thus, observing cavities and discarding that they are
product of the expansion of an H{\sc ii} region and/or winds from massive stars is important to investigate whether
collision between clouds could have happened \citep{torii15}.
Another observational signature of cloud-cloud collision is the bridge feature that can be seen in a position-velocity 
diagram from emission of molecular lines (see Fig.\,\ref{brid}). This feature probes the enhancement of the turbulent motion of 
the gas at the interface between both colliding clouds \citep{haworth15,fukui16,torii17}.

\subsection{Intrinsic processes}

After the above enumeration of some of the main triggered star formation mechanisms, this Section presents a few words 
concerning to some intrinsic processes related to the formation of a new star which have consequences in the 
surrounding ISM, such as jets and winds.

\begin{figure}[!h]
  \centering
  \includegraphics[width=0.47\textwidth]{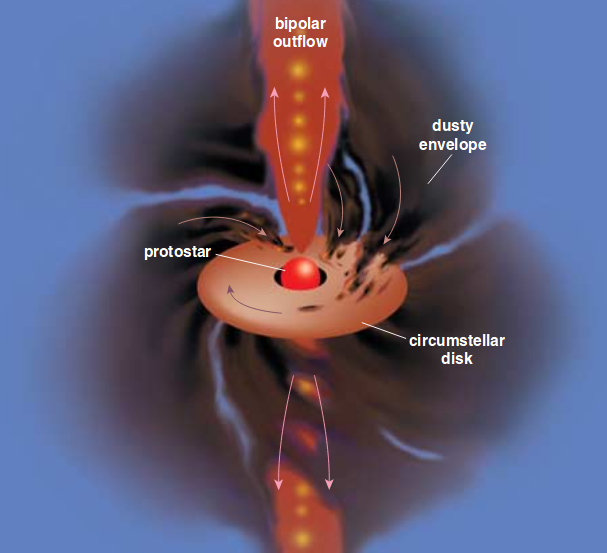}
  \caption{Typical sketch of the intrinsic processes ocurring during the formation of a star. Image from \citet{greene01}. }
  \label{form}
\end{figure}

Figure\,\ref{form} displays a typical sketch of the main star formation processes and features: an accretion disk that extracts matter
from the molecular envelope and feeds the central object, while powerful winds emanate from the poles of it,  
forming bipolar jets that sweep up molecular material generating the so called molecular outflows. 
There is a lot of literature concerning to these star forming processes, for example a recent review about protostellar
outflows can be found in \citet{bally16}, and for a more global treatment see \citet{mckee07}, who presented a framework to build a 
comprehensive theory of star formation, that as the authors mentioned, will be tested by the next generation of telescopes. 
After a decade of this review, I think that we are begining to do this work.

For instance, recently \citet{bje16} presented an important observational result: from molecular line observations towards a YSO in
TMC1A using the Atacama Submillimeter Telescope Array, it was demonstrated that the outflowing gas is launched from an extended 
region, and not from a region very close to the protostar as shown in Fig.\,\ref{form}. This result tends to prove that 
the outflowing activity may be due to winds from the disk. 

\begin{figure}[!h]
  \centering
  \includegraphics[width=0.47\textwidth]{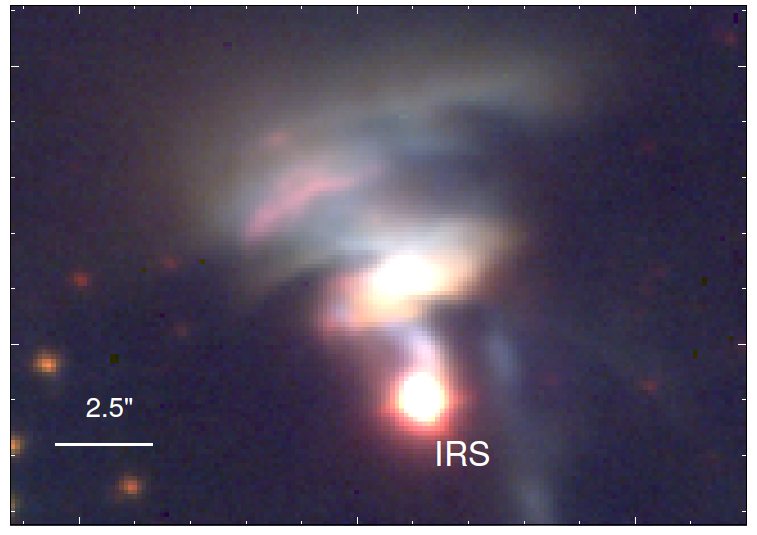}
  \caption{Example of circumstellar features at near-IR observed towards a jet precessing YSO. The angular resolution of this
image is about 0.5 arcsec, which taking into account the distance to the source, represents 400 AU. Image from \citet{paron16}. }
  \label{jet}
\end{figure}

Another observational results that changed the ideal picture of Fig.\,\ref{form}, is that the jets and outflows do not allways 
appears aligned.
Jets may present S- and Z-shaped symmetries, which indicate that the outflow axis has changed over time, 
probably due to precession induced by a companion, interactions with companion stars, or by the misalignement between the 
protostar rotation axis and magnetic fields \citep{bally07,lewis15}. Additionally, C-shaped morphology in jets may indicate 
motion of surrounding gas or motion of the outflow source itself. Figure\,\ref{jet} displays, with great detail, the typical features
observed at near-IR of a precessing jet: a cone-shaped nebula with some arc-like features. These structures are evidences of a cavity 
cleared in the circumstellar material that shines at near-IR due to radiation from the central protostar that is scattered at 
the inner walls of the cavity, emission from warm dust, and lines emission ([FeII], shock-excited H$_{2}$, between others).
Studying the outflow (a)symmetries is an important issue because it provides information about the dynamical environment
of the outflows engine and the interstellar medium in which they spread.

Finally, and related to the begining of this article, it is important to note that the stellar jets and outflows 
have deep implications in the chemistry of the cloud in which the stellar object is embedded. 
Either due to collisions between the jet and cloud, or the radiation generated by the central object and the jet itself, 
the chemistry can drastically change through ionization and/or evaporation of 
molecular species from the dust grains. Many molecules form at the interfaces between protostellar outflows and their 
natal molecular clouds \citep{van98,lim01,will02,codella06,roll14}, and their observation sheds light on such star 
forming process and its interaction with the medium.

\begin{acknowledgement}

I thank to the SOC and LOC of the 60 Meeting of the Asociación Argentina de Astronomía for the invitation to 
give a talk and to present this invited report in BAAA. I am grateful with Dr. M. Ortega for some useful comments
to this article.
This work was partially supported by grants awarded by CONICET, ANPCYT and UBA (UBACyT) from Argentina.

\end{acknowledgement}


\bibliographystyle{baaa}
\small
\bibliography{biblio}
 
\end{document}